# "Quark Mass Matrices with Four and Five Texture Zeroes, and the CKM Matrix, in terms of Mass Eigenvalues"

Bipin R. Desai and Alexander R. Vaucher

Department of Physics, University of California
Riverside, California 92521, USA

Abstract

Using the triangular matrix techniques of Kuo et al and Chiu et al for the four and five texture zero cases, with vanishing (11) elements for *U* and *D* matrices, it is shown, from the general eigenvalue equations and hierarchy conditions, that the quark mass matrices, and the CKM matrix can be expressed (except for the phases) entirely in terms of quark masses. The matrix structures are then quite simple and transparent. We confirm their results for the five texture zero case but find, upon closer examination of all the CKM elements which our results provide, that six of their nine patterns for the four texture zero case are not compatible with experiments. In total, only one five-texture zero and three four-texture zero patterns are allowed.

## **Introduction**

Quark Yukawa matrices in the upper triangular form provide a convenient tool to relate the CKM elements and quark masses as has been stressed recently by Kuo et al [1,2] and Chiu et al [3]. By incorporating the hierarchy of the physical parameters they show that the right-handed rotations can be selected appropriately, with small angles, to transform the Yukawa matrices into the triangular form. All the elements in the lower corner of this matrix are zero, in the hierarchical sense, and the rest are simply related to the products of quark masses and CKM angles.

Schematically, for a $3\times 3$ quark Yukawa matrix

$$Q=\begin{pmatrix} * & * & * \\ * & * & * \\ * & * & * \end{pmatrix}$$

the series of operations described above proceed as follows

$$QR=\begin{pmatrix} * & * & * \\ * & * & * \\ * & * & * \end{pmatrix}R=\begin{pmatrix} * & * & * \\ 0 & * & * \\ 0 & 0 & * \end{pmatrix}=T \qquad (1)$$

$$L^{+}QR=L^{+}T=L^{+}\begin{pmatrix} * & * & * \\ 0 & * & * \\ 0 & 0 & * \end{pmatrix}=\begin{pmatrix} m_1 & 0 & 0 \\ 0 & m_2 & 0 \\ 0 & 0 & m_3 \end{pmatrix}=Q^{diag} \qquad (2)$$

where the $m_i$'s are the quark mass eigenvalues. From (2) we note that the following matrix

$$T\,L^{+} = L\,Q^{diag}\,L^{+} \qquad (3)$$

is the symmetrized (Hermitian) version of *T*. It is this type of matrix one considers in order to discuss the texture zeros.

The CKM matrix is given by

$$V_{CKM} = L_u^{+}\,L_d \qquad (4)$$

where $L_u$, $L_d$ are the left-handed rotations for the up-quark (U) and down-quark (D) matrices, and where each $L$ is expressed as a product of three rotations,

$$L = L_{23}\,L_{13}\,L_{12}$$

In the small angle approximation,

$$L_{23} = \begin{pmatrix} 1 & 0 & 0 \\ 0 & 1 & s_{23} \\ 0 & -s_{23} & 1 \end{pmatrix}, \qquad s_{23} = \sin(\theta_{23})$$

similarly for the other two rotations.

From the triangular matrices it is found that

$$s_{ij} = \frac{T_{ij}}{T_{ii}} \qquad (5)$$

for both $T^u$ and $T^d$ ,where, actually, the diagonal elements $T_{ii}$, because of hierarchy, are the masses $m_i$ themselves. From (4) one can then obtain $V_{CKM}$.

The above results are quite useful, as discussed in ref. 1,2,3, if there are texture zeroes i.e. if the mass matrix is symmetric and one or more of its elements $Q_{ij}$ vanishes (at some scale).

We consider the heavy quark basis (i.e. (33) as the largest element). And we take at least one of the zeroes to correspond to the (11) diagonal element, i.e.

$$Q_{11} = 0$$

We find, remarkably, that all the matrix elements of Q and T are determined, apart from the phases, in terms of the mass eigenvalues $m_1, m_2, m_3$. As a result, **all** the CKM elements can be expressed in terms of these masses as well. Thus we are able to simplify the matrix structure beyond what was achieved in ref. 2,3. One of the consequences of our results is that some of the U-D combinations ("patterns") considered acceptable in ref.1,2,3 are found not to be so upon closer examination of the CKM elements.

We will discuss this in the following after a brief review of the triangular matrices.

**Triangular matrices and hierarchy**

For symmetric U and D matrices, a general hierarchical form is given as follows [1,2,3]

$$U = \begin{pmatrix} a_u\lambda^8 & b_u\lambda^6 & c_u\lambda^4 \\ b_u\lambda^6 & d_u\lambda^4 & e_u\lambda^2 \\ c_u\lambda^4 & e_u\lambda^2 & 1 \end{pmatrix} \qquad (6)$$

$$D = \begin{pmatrix} a_d\lambda^4 & b_d\lambda^3 & c_d\lambda^3 \\ b_d\lambda^3 & d_d\lambda^2 & e_d\lambda^2 \\ c_d\lambda^3 & e_d\lambda^2 & 1 \end{pmatrix} \qquad (7)$$

where $\lambda = |V_{us}| = 0.22$.

The hierarchy of the quark masses themselves is given by the diagonal elements of (6) and (7) above.

If we designate the triangular matrix as:

$$T = \begin{pmatrix} A & B & C \\ 0 & D & E \\ 0 & 0 & F \end{pmatrix} \tag{8}$$

then it is easy to confirm, on the basis of hierarchy in (6), (7), and (3), that the corresponding symmetric forms are given by [1,2,3],

$$U = \begin{pmatrix} X & Y & C \\ Y & Z & E \\ C & E & F \end{pmatrix} \tag{9}$$

where,

$$\begin{aligned} X &= A + \frac{B^2}{D} + \frac{C^2}{F} \\ Y &= B + \frac{CE}{F} \\ Z &= D + \frac{E^2}{F} \end{aligned} \tag{10}$$

and

$$D = \begin{pmatrix} X & B & C \\ B & D & E \\ C & E & F \end{pmatrix} \tag{11}$$

$$X = A + \frac{B^2}{D} \tag{12}$$

The matrix elements above are assumed to be real. The phases can be incorporated in the usual manner (with *A,D,F* real) as

$$B = |B| e^{i\phi_1} \; , \;\; C = |C| e^{i\phi_2} \; , \;\; E = |E| e^{i\phi_3}$$

with different $\phi_i$'s for the *U* and *D* systems. However, in the following, in view of the lack of firm experimental knowledge of the phases, we will be concentrating only on the absolute values of the CKM elements $\left|(V_{CKM})_{ij}\right|$, ignoring the phases. We will examine whether our predicted values fall within the experimental range.

**Texture zeroes, and U and D in terms of mass eigenvalues**

For a matrix Q whose diagonal form is given by

$$\begin{pmatrix} m_1 & 0 & 0 \\ 0 & m_2 & 0 \\ 0 & 0 & m_3 \end{pmatrix} \qquad (13)$$

the equation for the eigenvalues gives the following relations [4]

$$(1) + (2) + (3) = m_1 + m_2 + m_3 \qquad (14)$$

$$(12) + (13) + (23) = m_1 m_2 + m_1 m_3 + m_2 m_3 \qquad (15)$$

$$(123) = m_1 m_2 m_3 \qquad (16)$$

where, on the left hand side of the above equations,

- (*i*) represents the diagonal element $Q_{ii}$,
- (*ij*) represents the determinant of the (*ij*) sub-matrix of Q, and
- (123) represents the determinant of Q .

Applying the above conditions on U and D we note that with at least one texture zero at the (11) position, we have an additional condition

$$X=0 \quad . \tag{17}$$

We solve equations (14) through (17) simultaneously, along with other possible equations involving additional texture zero conditions, so that the solutions are consistent with the hierarchy limits indicated by (6) and (7). That is, we restrict our solutions not to exceed (but can be less than) the hierarchy bounds. In particular, we assume, as in ref 1,2,3, that the leading terms for the diagonal elements are given by

$$A = m_1 \ , \ D = m_2 \ , \ F = m_3 \tag{18}$$

We then find that all the matrix elements are determined in terms of $m_1$, $m_2$, and $m_3$, allowing for the re-phasing of the fermion masses. In particular from the triangular form (8) and relation (5) we determine all the CKM elements in terms of the masses.

In ref. 3 eight matrices were considered, $M_1 - M_8$, each with different locations for the texture zeros, in addition to $X = 0$. In Table I we have summarized our matrix structure for these eight matrices in terms of $m_1, m_2, m_3$ in their triangular forms $T^u$ and $T^d$. It is interesting to note that a structure exists for U for all the cases. However, that is not true for D, particularly when the texture zero occurs for the (22) element implying $m_2 = 0$ which is, of course, not allowed, or when the (12) element vanishes implying $m_1 = 0$.

In Table II we have given the combinations ("patterns") of U-D with a total of five texture zeros that were first considered by Ramond et al [5], and, recently by ref. 2,3. Since, as explained above, we now know the matrix structure of the T's in terms of the mass eigenvalues, the CKM elements can be obtained from (5). Three of the elements, $|V_{us}|$, $|V_{ub}|$, $|V_{cb}|$ are also given.

We reproduce all the results of ref. 2,3 and confirm that only pattern 3 is consistent with the experimental value of the $V_{CKM}$ elements [6] and the quark mass parameters (at $M_Z$) [7] [8].

In Table III we have given nine different patterns considered in ref. 3 with a total of four texture zeros (with one of the zeros at the (11) position). The corresponding CKM elements for each pattern are given as well. Those patterns for which all the CKM elements are consistent with the currently available experimental values have "OK" in the last column. Those that do not are indicated by "NO" and, in parenthesis, the CKM elements that are inconsistent are explicitly written.

It is found that patterns 1,3,4,7,8,9 with four texture zeroes are not compatible with the experimental values [6] [7] [8], contrary to the results of ref.3. And pattern 6 is questionable for $V_{cb}$. Because we now have a much fuller knowledge of the structure of the mass matrices, we have been able to predict all of the CKM elements and, consequently, find, upon closer examination, that many of the patterns of ref.3 are actually not consistent with experiments.

## Conclusion

The allowed patterns (including pattern 6) are given below for the five and four texture zeroes.

Five Texture Zeroes

$$(M_6, M_3) \quad U = \begin{pmatrix} 0 & 0 & \sqrt{m_u m_t} \\ 0 & m_c & 0 \\ \sqrt{m_u m_t} & 0 & m_t \end{pmatrix} \quad , \; D = \begin{pmatrix} 0 & \sqrt{m_d m_s} & 0 \\ \sqrt{m_d m_s} & m_s & \sqrt{m_d m_b} \\ 0 & \sqrt{m_d m_b} & m_b \end{pmatrix}$$

Four Texture Zeroes

$$(M_2, M_3) \quad U = \begin{pmatrix} 0 & 0 & \sqrt{m_u m_t} \\ 0 & m_c & \sqrt{m_u m_t} \\ \sqrt{m_u m_t} & \sqrt{m_u m_t} & m_t \end{pmatrix} \quad , \; D = \begin{pmatrix} 0 & \sqrt{m_d m_s} & 0 \\ \sqrt{m_d m_s} & m_s & \sqrt{m_d m_b} \\ 0 & \sqrt{m_d m_b} & m_b \end{pmatrix}$$

$$(M_4, M_3) \quad U = \begin{pmatrix} 0 & \sqrt{\frac{m_u m_c^2}{m_t}} & \sqrt{m_u m_t} \\ \sqrt{\frac{m_u m_c^2}{m_t}} & m_c & 0 \\ \sqrt{m_u m_t} & 0 & m_t \end{pmatrix} \quad , D = \begin{pmatrix} 0 & \sqrt{m_d m_s} & 0 \\ \sqrt{m_d m_s} & m_s & \sqrt{m_d m_b} \\ 0 & \sqrt{m_d m_b} & m_b \end{pmatrix}$$

$$(M_5, M_3) \quad U = \begin{pmatrix} 0 & \sqrt{m_u m_c} & \sqrt{m_u m_t} \\ \sqrt{m_u m_c} & 0 & \sqrt{m_c m_t} \\ \sqrt{m_u m_t} & \sqrt{m_c m_t} & m_t \end{pmatrix} \quad , \; D = \begin{pmatrix} 0 & \sqrt{m_d m_s} & 0 \\ \sqrt{m_d m_s} & m_s & \sqrt{m_d m_b} \\ 0 & \sqrt{m_d m_b} & m_b \end{pmatrix}$$

The five-texture zero pattern above and the first two four-texture zero patterns give a remarkably simple relation for the CKM elements, which are consistent with experiments [6],[7],

$$|V_{us}| \approx \sqrt{\frac{m_d}{m_s}}, \; |V_{ub}| \approx \sqrt{\frac{m_u}{m_t}}, \; |V_{cb}| \approx \sqrt{\frac{m_d}{m_b}}.$$

Interestingly, as already pointed out in ref.3, the currently most popular pattern

$$(M_3, M_3) \quad U = \begin{pmatrix} 0 & \sqrt{m_u m_c} & 0 \\ \sqrt{m_u m_c} & m_c & \sqrt{m_u m_t} \\ 0 & \sqrt{m_u m_t} & m_t \end{pmatrix} \quad , D = \begin{pmatrix} 0 & \sqrt{m_d m_s} & 0 \\ \sqrt{m_d m_s} & m_s & \sqrt{m_d m_b} \\ 0 & \sqrt{m_d m_b} & m_b \end{pmatrix}$$

is NOT among the allowed patterns.

We re-iterate that the advantage our results have is that the mass and CKM matrices are quite transparent and simple as they are expressed in terms of just the mass elements. All matrix elements are, therefore, determined (except for the phase). A comparison with experiments reveals that for those mass-matrices with a vanishing (11) element, only one five-texture zero and three four-texture zero structures exist.

This work was supported in part by the U.S. Department of Energy under contract NO. DE-F603-94ER40837.

Table I

| $M_i$ | Texture Zeroes | $T^u$ | $T^d$ |
|---|---|---|---|
| $M_1$ | $\begin{pmatrix} 0 & * & * \\ * & * & * \\ * & * & * \end{pmatrix}$ | $\begin{pmatrix} m_1 & \sqrt{\frac{m_1 m_2^2}{m_3}} & \sqrt{m_1 m_3} \\ 0 & m_2 & \sqrt{m_1 m_3} \\ 0 & 0 & m_3 \end{pmatrix}$ | $\begin{pmatrix} m_1 & \sqrt{m_1 m_2} & 2\sqrt{\frac{m_1^2 m_3}{m_2}} \\ 0 & m_2 & \sqrt{m_1 m_3} \\ 0 & 0 & m_3 \end{pmatrix}$ |
| $M_2$ | $\begin{pmatrix} 0 & 0 & * \\ 0 & * & * \\ * & * & * \end{pmatrix}$ | $\begin{pmatrix} m_1 & m_1 & \sqrt{m_1 m_3} \\ 0 & m_2 & \sqrt{m_1 m_3} \\ 0 & 0 & m_3 \end{pmatrix}$ | ------------------- |
| $M_3$ | $\begin{pmatrix} 0 & * & 0 \\ * & * & * \\ 0 & * & * \end{pmatrix}$ | $\begin{pmatrix} m_1 & \sqrt{m_1 m_2} & 0 \\ 0 & m_2 & \sqrt{m_1 m_3} \\ 0 & 0 & m_3 \end{pmatrix}$ | $\begin{pmatrix} m_1 & \sqrt{m_1 m_2} & 0 \\ 0 & m_2 & \sqrt{m_1 m_3} \\ 0 & 0 & m_3 \end{pmatrix}$ |
| $M_4$ | $\begin{pmatrix} 0 & * & * \\ * & * & 0 \\ * & 0 & * \end{pmatrix}$ | $\begin{pmatrix} m_1 & \sqrt{\frac{m_1 m_2^2}{m_3}} & \sqrt{m_1 m_3} \\ 0 & m_2 & 0 \\ 0 & 0 & m_3 \end{pmatrix}$ | $\begin{pmatrix} m_1 & \sqrt{m_1 m_2} & \sqrt{\frac{m_1 m_2}{2}} \\ 0 & m_2 & 0 \\ 0 & 0 & m_3 \end{pmatrix}$ |
| $M_5$ | $\begin{pmatrix} 0 & * & * \\ * & 0 & * \\ * & * & * \end{pmatrix}$ | $\begin{pmatrix} m_1 & \sqrt{\frac{2 m_1 m_2^2}{m_3}} & \sqrt{m_1 m_3} \\ 0 & m_2 & \sqrt{m_2 m_3} \\ 0 & 0 & m_3 \end{pmatrix}$ | ------------------- |
| $M_6$ | $\begin{pmatrix} 0 & 0 & * \\ 0 & * & 0 \\ * & 0 & * \end{pmatrix}$ | $\begin{pmatrix} m_1 & 0 & \sqrt{m_1 m_3} \\ 0 & m_2 & 0 \\ 0 & 0 & m_3 \end{pmatrix}$ | ------------------- |
| $M_7$ | $\begin{pmatrix} 0 & * & 0 \\ * & * & 0 \\ 0 & 0 & * \end{pmatrix}$ | $\begin{pmatrix} m_1 & \sqrt{m_1 m_2} & 0 \\ 0 & m_2 & 0 \\ 0 & 0 & m_3 \end{pmatrix}$ | $\begin{pmatrix} m_1 & \sqrt{m_1 m_2} & 0 \\ 0 & m_2 & 0 \\ 0 & 0 & m_3 \end{pmatrix}$ |
| $M_8$ | $\begin{pmatrix} 0 & * & 0 \\ * & 0 & * \\ 0 & * & * \end{pmatrix}$ | $\begin{pmatrix} m_1 & \sqrt{m_1 m_2} & 0 \\ 0 & m_2 & \sqrt{m_2 m_3} \\ 0 & 0 & m_3 \end{pmatrix}$ | ------------------- |

Table II: Five Texture Zero Patterns. ("*Exp.*" indicates experimental value).

| Pattern | $U$ | $D$ | $\lvert V_{us}\rvert$ (*Exp.* 0.22) | $\lvert V_{ub}\rvert$ (*Exp.* 0.0036) | $\lvert V_{cb}\rvert$ (*Exp.* 0.040) | |
|---|---|---|---|---|---|---|
| 1 $M_7, M_3$ | $\begin{pmatrix} 0 & * & 0 \\ * & * & 0 \\ 0 & 0 & * \end{pmatrix}$ | $\begin{pmatrix} 0 & * & 0 \\ * & * & * \\ 0 & * & * \end{pmatrix}$ | $\sqrt{\frac{m_d}{m_s}} \pm \sqrt{\frac{m_u}{m_c}}$ (0.17, 0.28) | $\sqrt{\frac{m_d m_u}{m_b m_c}}$ 0.0023 | $\sqrt{\frac{m_d}{m_b}}$ 0.040 | No ($V_{ub}$) |
| 2 $M_8, M_3$ | $\begin{pmatrix} 0 & * & 0 \\ * & 0 & * \\ 0 & * & * \end{pmatrix}$ | $\begin{pmatrix} 0 & * & 0 \\ * & * & * \\ 0 & * & * \end{pmatrix}$ | $\sqrt{\frac{m_d}{m_s}} \pm \sqrt{\frac{m_u}{m_c}}$ (0.17,0.28) | $\sqrt{\frac{m_u}{m_c}}\left[\sqrt{\frac{m_c}{m_t}} \pm \sqrt{\frac{m_d}{m_b}}\right]$ (0.0011, 0.0058) | $\sqrt{\frac{m_c}{m_t}} \pm \sqrt{\frac{m_d}{m_b}}$ (0.022, 0.10) | No ($V_{ub}, V_{cb}$) |
| 3 $M_6, M_3$ | $\begin{pmatrix} 0 & 0 & * \\ 0 & * & 0 \\ * & 0 & * \end{pmatrix}$ | $\begin{pmatrix} 0 & * & 0 \\ * & * & * \\ 0 & * & * \end{pmatrix}$ | $\sqrt{\frac{m_d}{m_s}}$ 0.22 | $\sqrt{\frac{m_u}{m_t}}$ 0.0036 | $\sqrt{\frac{m_d}{m_b}}$ 0.040 | OK |
| 4 $M_3, M_7$ | $\begin{pmatrix} 0 & * & 0 \\ * & * & * \\ 0 & * & * \end{pmatrix}$ | $\begin{pmatrix} 0 & * & 0 \\ * & * & 0 \\ 0 & 0 & * \end{pmatrix}$ | $\sqrt{\frac{m_d}{m_s}} \pm \sqrt{\frac{m_u}{m_c}}$ (0.17,0.28) | $\sqrt{\frac{m_u^2}{m_c m_t}}$ 0.00021 | $\sqrt{\frac{m_u}{m_t}}$ 0.0036 | No ($V_{ub}, V_{cb}$) |
| 5 $M_2, M_7$ | $\begin{pmatrix} 0 & 0 & * \\ 0 & * & * \\ * & * & * \end{pmatrix}$ | $\begin{pmatrix} 0 & * & 0 \\ * & * & 0 \\ 0 & 0 & * \end{pmatrix}$ | $\sqrt{\frac{m_d}{m_s}} \pm \frac{m_u}{m_c}$ (0.22,0.23) | $\sqrt{\frac{m_u}{m_t}}$ 0.0036 | $\sqrt{\frac{m_u}{m_t}}$ 0.0036 | No ($V_{cb}$) |

Table III: Four Texture Zero Patterns. ("*Exp.*" indicates experimental value).

| Pattern | $U$ | $D$ | $\lvert V_{us}\rvert$ (*Exp.* 0.22) | $\lvert V_{ub}\rvert$ (*Exp.* 0.0036) | $\lvert V_{cb}\rvert$ (*Exp.* 0.040) | |
|---|---|---|---|---|---|---|
| 1 $M_1, M_7$ | $\begin{pmatrix} 0 & * & * \\ * & * & * \\ * & * & * \end{pmatrix}$ | $\begin{pmatrix} 0 & * & 0 \\ * & * & 0 \\ 0 & 0 & * \end{pmatrix}$ | $\sqrt{\frac{m_d}{m_s}} \pm \sqrt{\frac{m_u}{m_c}}$ (0.17,0.28) | $\sqrt{\frac{m_u}{m_t}}$ 0.0036 | $\sqrt{\frac{m_u}{m_t}}$ 0.0036 | No ($V_{cb}$) |
| 2 $M_2, M_3$ | $\begin{pmatrix} 0 & 0 & * \\ 0 & * & * \\ * & * & * \end{pmatrix}$ | $\begin{pmatrix} 0 & * & 0 \\ * & * & * \\ 0 & * & * \end{pmatrix}$ | $\sqrt{\frac{m_d}{m_s}}$ 0.22 | $\sqrt{\frac{m_u}{m_t}}$ 0.0036 | $\sqrt{\frac{m_d}{m_b}} \pm \sqrt{\frac{m_u}{m_t}}$ (0.036,0.043) | OK |
| 3 $M_2, M_4$ | $\begin{pmatrix} 0 & 0 & * \\ 0 & * & * \\ * & * & * \end{pmatrix}$ | $\begin{pmatrix} 0 & * & * \\ * & * & 0 \\ * & 0 & * \end{pmatrix}$ | $\sqrt{\frac{m_d}{m_s}}$ 0.22 | $\sqrt{\frac{m_d m_s}{2m_b^2}} \pm \sqrt{\frac{m_u}{m_t}}$ (0.0013,0.0085) | $\sqrt{\frac{m_u}{m_t}}$ 0.0036 | No ($V_{cb}$) |
| 4 $M_3, M_4$ | $\begin{pmatrix} 0 & * & 0 \\ * & * & * \\ 0 & * & * \end{pmatrix}$ | $\begin{pmatrix} 0 & * & * \\ * & * & 0 \\ * & 0 & * \end{pmatrix}$ | $\sqrt{\frac{m_d}{m_s}} \pm \sqrt{\frac{m_u}{m_c}}$ (0.17,0.28) | $\sqrt{\frac{m_d m_s}{2m_b^2}} \pm \sqrt{\frac{m_u^2}{m_c m_t}}$ (0.0047, 0.0051) | $\sqrt{\frac{m_u}{m_t}}$ 0.0036 | No ($V_{ub}$, $V_{cb}$) |
| 5 $M_4, M_3$ | $\begin{pmatrix} 0 & * & * \\ * & * & 0 \\ * & 0 & * \end{pmatrix}$ | $\begin{pmatrix} 0 & * & 0 \\ * & * & * \\ 0 & * & * \end{pmatrix}$ | $\sqrt{\frac{m_d}{m_s}} \pm \sqrt{\frac{m_u}{m_t}}$ (0.22,0.23) | $\sqrt{\frac{m_u}{m_t}}$ 0.0036 | $\sqrt{\frac{m_d}{m_b}}$ 0.040 | OK |
| 6 $M_5, M_3$ | $\begin{pmatrix} 0 & * & * \\ * & 0 & * \\ * & * & * \end{pmatrix}$ | $\begin{pmatrix} 0 & * & 0 \\ * & * & * \\ 0 & * & * \end{pmatrix}$ | $\sqrt{\frac{m_d}{m_s}} \pm \sqrt{\frac{2m_u}{m_t}}$ (0.22,0.23) | $\sqrt{\frac{m_u}{m_t}}$ 0.0036 | $\sqrt{\frac{m_c}{m_t}} \pm \sqrt{\frac{m_d}{m_b}}$ (0.022,0.10) | ? ($V_{cb}$) |
| 7 $M_6, M_1$ | $\begin{pmatrix} 0 & 0 & * \\ 0 & * & 0 \\ * & 0 & * \end{pmatrix}$ | $\begin{pmatrix} 0 & * & * \\ * & * & * \\ * & * & * \end{pmatrix}$ | $\sqrt{\frac{m_d}{m_s}}$ 0.22 | $\sqrt{\frac{m_u}{m_t}} \pm 2\sqrt{\frac{m_d^{\ 2}}{m_s m_b}}$ (0.014,0.021) | $\sqrt{\frac{m_d}{m_b}}$ 0.040 | No ($V_{ub}$) |
| 8 $M_7, M_1$ | $\begin{pmatrix} 0 & * & 0 \\ * & * & 0 \\ 0 & 0 & * \end{pmatrix}$ | $\begin{pmatrix} 0 & * & * \\ * & * & * \\ * & * & * \end{pmatrix}$ | $\sqrt{\frac{m_d}{m_s}} \pm \sqrt{\frac{m_u}{m_c}}$ (0.17,0.28) | $2\sqrt{\frac{m_d^{\ 2}}{m_s m_b}} \pm \sqrt{\frac{m_d m_u}{m_b m_c}}$ (0.015,0.020) | $\sqrt{\frac{m_d}{m_b}}$ 0.040 | No ($V_{ub}$) |
| 9 $M_8, M_1$ | $\begin{pmatrix} 0 & * & 0 \\ * & 0 & * \\ 0 & * & * \end{pmatrix}$ | $\begin{pmatrix} 0 & * & * \\ * & * & * \\ * & * & * \end{pmatrix}$ | $\sqrt{\frac{m_d}{m_s}} \pm \sqrt{\frac{m_u}{m_c}}$ (0.17,0.28) | $2\sqrt{\frac{m_d^{\ 2}}{m_s m_b}} \pm \sqrt{\frac{m_d m_u}{m_b m_c}}$ (0.015,0.020) | $\sqrt{\frac{m_c}{m_t}} \pm \sqrt{\frac{m_d}{m_b}}$ (0.022,0.10) | No ($V_{ub}$, $V_{cb}$) |